\documentclass[pss]{wiley2sp} 
\usepackage{amsmath}

\begin{document}


\title{Electromagnetic absorption of quasi-1D Majorana nanowires}

\titlerunning{EM absorption of quasi-1D Majorana nanowires}

\author{%
  Javier Osca\textsuperscript{\Ast,\textsf{\bfseries 1}},
  Lloren\c{c} Serra \textsuperscript{\textsf{\bfseries 1,2}}
  }

\authorrunning{Osca and Serra}

\mail{e-mail
  \textsf{javier.osca@uib.es}, Phone:
  +34-971259880, Fax: +34-971173248}

\institute{%
  \textsuperscript{1}\,
  Institut de F\'{\i}sica Interdisciplin\`aria i de Sistemes Complexos IFISC (CSIC-UIB), 
  E-07122 Palma de Mallorca, Spain\\
  \textsuperscript{2}\,
  Departament de F\'{\i}sica, Universitat de les Illes Balears, E-07122 Palma de Mallorca, Spain}

\received{XXXX, revised XXXX, accepted XXXX} 
\published{XXXX} 


\abstract{%
%
%
%
\abstcol{%
We calculate the electromagnetic absorption cross section of long and narrow 
nanowires, in the so-called quasi-1D limit. We consider only two transverse bands
and compute the dipole absorption cross section taking into account quasiparticle
transitions
from negative to positive energy eigenstates of the Bogoliubov-de Gennes Hamiltonian.
The presence of the zero energy (Majorana) state manifests in the 
different  absorption spectra for $x$ (parallel) and $y$ (transverse) polarizations
of the electromagnetic field.
}
{
In the $y$-polarized case, the Majorana state causes a low energy absorption plateau
extending from mid-gap up to  
full-gap energy. Increasing further the energy, the plateau is followed by a region 
of enhanced absorption due to transitions across the full gap.
For $x$ polarization the low energy absorption plateau is not observed.}
 }

%
%

\maketitle   

\section{Introduction}
The physics of Majorana states in semiconductor nanowires has been attracting much attention
in recent years \cite{Alicea,Beenakker,StanescuREV,Franz}. 
There is a fundamental motivation, as they are a novel realization of the 
physics envisioned by Majorana   for a class of elementary particles 
already in 1937 \cite{Majorana}.
Majorana states in semiconductor nanowires are also interesting from the point of view of technological applications. Indeed, it has been suggested that their character of nonabelian
anyons with topological protection could be exploited for implementing quantum computation in 
practice \cite{Nayak}.

Several  electrical transport experiments with semiconductor nanowires have observed a zero bias peak for a certain range of magnetic fields,
consistent with  a zero energy Majorana state \cite{Mourik,Deng,Das,Finck}. 
The observed peak height is, however, an order of magnitude lower than the quantized value $2e^2/h$. This discrepancy is not yet well understood, as it might be due to effects ranging from finite temperature, experimental and tunneling resolutions to other low-energy subgap states and possible inelastic and renormalization processes \cite{Pen,Sar_prep}.
It is, therefore, important to 
characterize the Majorana mode and its role in different experimental signals 
in order to ascertain the existence of such peculiar states. 

\begin{figure}[b]%
  \includegraphics*[width=.4\textwidth]{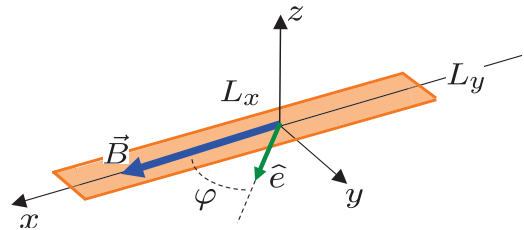}%
  \caption[]{%
Sketch of the system. }
    \label{F1_0}
\end{figure}

In Ref.\ \cite{Ruiz} we investigated the role of the Majorana state on the optical absorption 
of 2D semiconductor nanowires. It was suggested that 
the Majorana causes 
a low energy absorption feature, 
observed when the polarization of the electromagnetic (EM) field is in transverse direction to the nanowire (Fig.\ \ref{F1_0}).
We used a grid approach, 
whereby the $xy$ plane was discretized in a mesh of points and the Bogoliubov-de Gennes
eigenvalue  equation was transformed into a matrix diagonalization problem using 
finite differences on the grid. A shortcoming of that method is the high computational 
cost for fine grids, that aggravates when a large number of Hamiltonian
eigenstates need to be obtained. Indeed, the absorption cross section is the result 
of many quasiparticle transitions from occupied to empty eigenstates (see Fig.\ \ref{F1}).

In this work we extend the analysis of Ref.\ \cite{Ruiz} by focussing on the 
quasi-1D limit of very long and narrow nanowires. We use a mixed grid-basis approach,
discretizing the longitudinal $x$ coordinate in a grid and describing the transverse $y$ degree
of freedom in a basis of square well eigenstates (Fig. \ref{F1_0} contains the axis definitions).
As we are  interested in the case of narrow 
wires, we will restrict to only two transverse states.  
This technique is computationally 
less demanding  and it will allow us the calculation of large numbers of eigenstates,  
thus better characterizing the absorption cross section. 

\section{Model and linear absorption formalism}

\begin{figure}[t]%
  \includegraphics*[width=.45\textwidth]{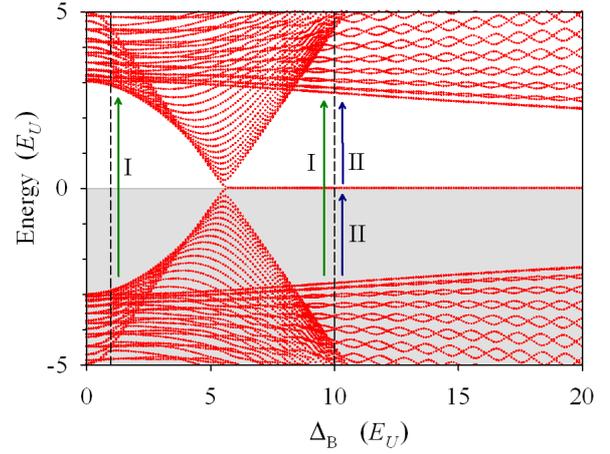}%
  \caption[]{%
 Evolution of the quasiparticle energy levels with the Zeeman energy.
 A zero energy mode is seen for $\Delta_B>5 E_U$. The vertical dashed lines 
 signal two values of $\Delta_B$ for which the cross section is shown in Fig.\ \ref{F2}.
 The shaded region is the Dirac sea of occupied quasiparticles with negative energies.  
 The arrows indicate the transitions across the full gap (I) and across the mid gap (II).
 The latter are possible only in presence of the Majorana state.
 Parameters: $\Delta_0=3\,E_U$, $\alpha=3.1\,E_UL_U$, $\mu=20\, E_U$, $L_y=0.5\,L_U$, 
 $L_x=30\, L_U$. }
    \label{F1}
\end{figure}

We use the model and notation already described in Ref.\ \cite{Ruiz}, to which the reader
is addressed for the details. The Hamiltonian reads 
\begin{eqnarray}
\label{eq1}
\mathcal{H}_{BdG}
&=&
\left[\frac{p^2_x+p^2_y}{2m}+V(x,y)-\mu\right]\tau_z + \Delta_B\, \sigma_x\nonumber\\
&+& \Delta_0\, \tau_x+\frac{\alpha}{\hbar}(p_x\sigma_y-p_y \sigma_x) \tau_z\; ,
\end{eqnarray}
where the potential $V(x,y)$ is a hard wall confinement to a rectangle of lengths 
$L_x$ and $L_y$, with $L_y<<L_x$. 
Equation (\ref{eq1}) contains the three mechanisms giving Majorana physics with semiconductor 
nanowires. Namely, these are the Zeeman ($\Delta_B$), superconductivity ($\Delta_0$) and Rashba 
spin-orbit ($\alpha$) interactions. 
As a consequence of particle-hole symmetry 
the eigenstates of ${\cal H}_{BdG}$ always come in pairs 
of energy $\pm E_i$, with $i=1,2,\dots$. When a critical value of the Zeeman parameter
is surpassed the system presents a pair of solutions at very small energies $E_1=\pm\epsilon$,
which is signaling the Majorana phase transition (Fig.\ \ref{F1}).

We present the results in effective length ($L_U$) and energy ($E_U=\hbar^2/mL_U^2$) units. For definiteness, 
we take $L_U=150\, {\rm nm}$ and $E_U=0.1\, {\rm meV}$. We also assume
$m=0.033\,m_e$ and $\alpha=47\, {\rm meV}{\rm nm}$, typical values for InAs nanowires. 
With a gyromagnetic factor of  $g=15$ the maximum Zeeman energy in Fig.\ \ref{F1} ($20\,E_U$)
corresponds to $\approx 4.4\,{\rm T}$.

The dipole absorption cross section, in the limit of weak EM field, is given by the 
eigenstate transitions,
\begin{equation}
\label{eq2}
\sigma(\omega)\approx  \sum_{k,s} 
\, \frac{\left|  
\langle k\vert \vec{p}\cdot \hat{e}\vert s\rangle
\right|^2}{\omega_{ks}}\,\delta(\omega-\omega_{ks})\, f_s\,(1-f_k)\; ,
\end{equation}
where $f_{s,k}$ are the occupations of levels $s,k$Á as given by Fermi functions
with a given temperature $T$, and $\hat{e}$ is a unitary vector giving  the polarization 
direction of the EM field. Below, we will restrict for simplicity to the limit of vanishing temperature
where only transitions from negative to positive energy eigenstates contribute 
to Eq.\ (\ref{eq2}), as sketched in Fig.\ \ref{F1}.

The eigenstates of ${\cal H}_{BdG}$  are obtained in a mixed representation, using a space grid
in $x$ and a set of square well eigenstates $\phi_n(y)$ in the transverse direction,
\begin{equation}
\label{eq3}
\Psi \equiv
\sum_{n=1,2,\,s_\sigma,s_\tau=\pm}{\!\!\!\!\!\!\!\!\!\!
\psi_{n s_\sigma s_\tau}(x)\, 
\phi_n(y)\,
\chi_{s_\sigma}(\eta_\sigma)\,
\chi_{s_\tau}(\eta_\tau)
}\; .
\end{equation}
The unknown functions $\psi_{n s_\sigma s_\tau}(x)$ are found on the grid
by matrix diagonalization. We are interested in the limit of narrow wires and will thus restrict to only 
two transverse modes $n=1,2$. This is the minimum needed in order to account for the possibility 
of dipole excitations in transverse polarization, i.e., with a $p_y$ operator
in Eq.\ (\ref{eq2}).

\section{Results}

Figure \ref{F1} displays the energy eigenstates as a function of the Zeeman parameter for a selected case.
The Majorana phase transition is clearly seen.
For vanishing field the spectrum has a gap of $\approx 2\Delta_0$ ($6\, E_U$); this is 
followed by an intermediate region without a clear gap and, at large-enough fields, the gap is approximately restored with the
qualitative difference of the Majorana pair of eigenstates lying right in the middle of the main gap. 
Of this particular pair, one state is infinitesimally below zero while the other is infinitesimally above.
We may thus expect mid-gap transitions, labelled as II in Fig.\ \ref{F1}. We notice here that dipole transitions between the two states of a given pair are fobidden because of particle-hole symmetry
\cite{Ruiz}.
 
\begin{figure*}[htb]%
  \sidecaption
  \includegraphics*[width=.63\textwidth]{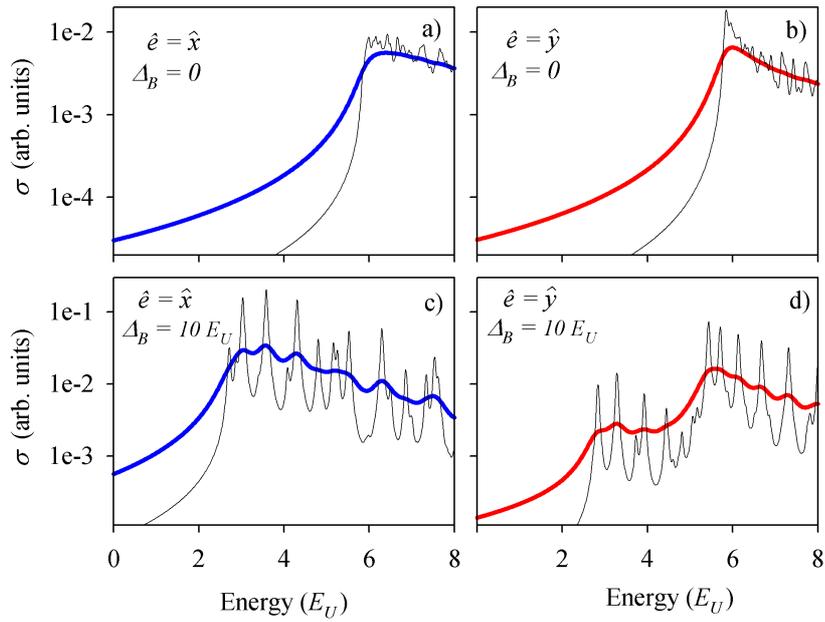}%
  \caption[]{%
Absorption cross sections for the spectra of Fig.\ \ref{F1}, corresponding to the 
Zeeman values indicated by dashed lines in that figure. Left and right panels are for $\hat{x}$
and $\hat{y}$ polarizations, repectively. In each panel, thin lines resolve individual peaks while
thick lines are the smoothened absorptions obtained with larger averaging widths of the Lorentzian peaks.
A lower absorption plateau starting around mid gap energy ($\approx 3\, E_U$), caused
by the Majorana state, is clearly seen in 
panel d).
}
\label{F2}
\end{figure*}

The dipole cross sections corresponding to the eigenstates of Fig.\ \ref{F1}
for two selected values of $\Delta_B$ 
are shown in Fig.\ \ref{F2}. The delta peaks of Eq.\ (\ref{eq2}) have been replaced by 
Lorentzian functions of width $\Gamma$ that could represent the experimental resolution
of the apparatus. We show both the result for a low $\Gamma$, resolving the individual transitions,
and for a high $\Gamma$, yielding a smoothened absorption profile. 
As expected, for vanishing magnetic field absorption occurs only above the full gap ($\approx 6\,E_U$)
and there is no significant difference between $x$ and $y$ polarizations (upper panels). 

The emergence of the Majorana state causes two remarkable modifications (lower panels in Fig\ \ref{F2}).
First, absorption starts at mid-gap energy $\approx 3\, E_U$ due to transitions of type II 
from and to the Majorana pair of states. Second, there is a qualitative difference between $x$ and
$y$ polarizations. For $x$ polarization there is a rather featureless smooth absorption
once the mid-gap threshold is overcome. For $y$ polarization, however, there is a lower absorption plateau 
extending from mid-gap to full-gap energies, followed by an enhanced absorption once the
full-gap energy is exceeded. 
The differences between $x$ and $y$ polarization in presence of a Majorana state were already 
suggested in Ref.\ \cite{Ruiz}. However, in that work the number of eigenstates was truncated
to lower values and thus the absorption for higher energies was less converged than in this work.

\section{Conclusions} We have improved the analysis of Ref.\ \cite{Ruiz} of the EM absorption 
of Majorana wires in the quasi-1D limit by considering larger sets of eigenstates.
We confirm that  the differences between $x$ and
$y$ polarized absorptions are an important signature of the presence of the Majorana state.    
In particular, with $y$ polarization the Majorana mode causes a low-energy absorption plateau, from mid-gap to 
full-gap energies, followed by an enhanced absorption once the full-gap energy is exceeded.
The present method can be easily extended to consider the effect of optical masks covering parts of the nanowire
or other quasi-1D geometries like L-junctions.

\begin{acknowledgement}
This work was funded by MINECO-Spain (grant FIS2011-23526),
CAIB-Spain (Conselleria d'Educaci\'o, Cultura i Universitats) and 
FEDER. We hereby acknowledge the PhD grant provided by the University 
of the Balearic Islands.
\end{acknowledgement}

%
\bibliographystyle{pss}
\bibliography{Articulo_4L}
%

\end{document}